\documentstyle[preprint,prb,aps]{revtex}
\begin{document}
\draft
\title{Mean-field phase diagrams of $AT_2X_2$ compounds}
\author{M.L. Plumer}
\address{ Centre de Recherche en Physique du Solide et D\'epartement de
Physique}
\address{Universit\'e de Sherbrooke, Sherbrooke, Qu\'ebec, Canada J1K 2R1}
\date{June 1994}
\maketitle
\begin{abstract}
Magnetic-field -- temperature phase diagrams of the
axial next-nearest-neighbor Ising model are calculated within
the framework of a Landau-type expansion of the free energy derived
from molecular-field theory.  Good qualitative agreement is found with recently
reported results on body-centered-tetragonal
$UPd_2Si_2$.  This work is expected to also be relevant for
related compounds.
\end{abstract}

There exists a large class of compounds with the body-centered-tetragonal
(bct) structure of the generic formula $AT_2X_2$, where $A$ represents either
$U$ or a rare-earth element, $T$ is a transition metal, and $X$ is either $Si$
or $Ge$.\cite{mail}  The materials of interest here have strong $c$-axial
anisotropy, with the possibility of long-range Ruderman-Kittel-Kasuya-Yosida
(RKKY) interactions.  It may thus be expected that the
axial next-nearest-neighbor Ising (ANNNI) model and its extensions
are relevant in cases where in-plane interactions are
ferromagnetic.\cite{selke}  It appears to be an empirical rule that
such models yield a succession of {\it principal} phases characterized by
increasing periodicities as the temperature is increased.  A counter example is
found in the compound $UNi_2Si_2$, where a more complicated model was
required to explain its observed sequence of magnetically ordered
states (Ref.\onlinecite{mail}, hereafter referred to as I).  Mean-field theory
proved inadequate in this instance.  All other relevant $AT_2X_2$ compounds,
however, appear to exhibit magnetic structures consistent with expectations
based a mean-field approximation (see I).  In the present work, the
magnetic-field -- temperature phase diagram of the ANNNI model is considered
within the framework of a Landau-type free energy developed from
molecular-field theory.  Good qualitative agreement is found with recently
reported results on $UPd_2Si_2$.\cite{shem,collins}

Magnetic interactions are assumed to be well described by a Heisenberg-type
Hamiltonian of the form
\begin{equation}
{\cal H}= - \frac12 \sum_{ij} J({\bf r}_i - {\bf r}_j) {\bf s}({\bf r}_i)
\cdot {\bf s}({\bf r}_j) - {\bf H} \cdot \sum_i {\bf s}({\bf r}_i),
\end{equation}
where the spin density ${\bf s}({\bf r}_i)$, as well as the applied
magnetic field ${\bf H}$, are assumed to lie along the bct $c$-axis
$(\| {\bf \hat z})$.  With ferromagnetic in-plane interactions $J_0 > 0$,
the problem effectively becomes one-dimensional within mean-field
theory.  It is the magnetic ordering between the ferromagnetic planes
separated by $c'=\frac12 c$ which is of interest.
A Landau-type free energy functional may be derived
from molecular-field theory, based on the analysis of Bak and von
Boehm\cite{bak} (also see references in I), with results to sixth-order
written as
\begin{eqnarray}
F &=& (1/2V) \int dz dz' A(\tau) s(z) s(z')
+ (B/4V) \int dz [s(z)]^4
\nonumber \\
&+& (C/6V) \int dz [s(z)]^6 + \cdot \cdot \cdot
- H \int dz s(z),
\end{eqnarray}
where $\tau = z - z'$ and
\begin{equation}
A(z)= aT\delta(z) - j^2 J(z)/V,
\end{equation}
$B=bT$, $C=cT$, with the coefficients $a,b,c$ given in terms of the
angular momentum $j$ through the Brillouin function (see I).  For simplicity,
we take here the classical limit $j \rightarrow \infty$ (with energies divided
by $j^2$) so that $a=3, b=\frac95, c \simeq 1.697$.

The ANNNI model with antiferromagnetic
next-nearest-neighbor coupling, $J_2 < 0$, is frustrated.
Numerous higher-order commensurate phases, as well an incommensurate state
(IC), result from mean-field analyses of a model with near-neighbor
interactions also antiferromagnetic, $J_1 < 0$, and with an additional
small ferromagnetic third-neighbor coupling, $J_3 > 0$.\cite{yam}
Monte-Carlo simulations seem to suggest, however, that only a few
princicpal commensurate phases survive the effects of critical
fluctuations.\cite{selk,hass}   For values of $J_2/J_1$ not too
large, only two commensurate states appear: The period-2
$\langle 1 \rangle$ state and the period-3 $\langle 12 \rangle$ state (see
Ref.\onlinecite{selke} for an explanation of the notation).  It is precisely
these two commensurate states which occur in the axial $AT_2X_2$ compounds
(see I).

With the limited goal of describing these two commensurate, as well as the IC,
states, the spin density can be written as
\begin{equation}
s(z) = m + Se^{iQz} + S^\ast e^{-iQz},
\end{equation}
where $m$ is the uniform component, $S$ is the complex polarization
vector, and $Q$ is the wave vector.  Using this expression in (2)
yields many types of Umklapp terms, non-zero only for $nQ=G$, where
$G$ is a reciprocal lattice vector.  For the purposes of the this
work, contributions with $n=1$ and $n=5$ are omitted.  It is instructive to
present the many terms which are relevant, and to write these as
\begin{equation}
F = F_2 + F_4 + F_6 - mH,
\end{equation}
with
\begin{equation}
F_2 = {\textstyle \frac12} A_0 m^2 + A_Q \mid S \mid ^2
+ {\textstyle \frac12} A_Q [S^2 + c.c.] \Delta_{2Q,G},
\end{equation}
\begin{eqnarray}
F_4 &=& {\textstyle \frac14} B \lbrace S_T^4
+ 2 \mid S^2 \mid ^2 + 8 \mid m S \mid ^2
\nonumber \\
&+& 2[ 2(mS)^2 +  S_T^2 S^2 + c.c.] \Delta_{2Q,G}
\nonumber \\
&+& 4[mS^3 + c.c.] \Delta_{3Q,G} + [ S^4 + c.c.] \Delta_{4Q,G} \rbrace,
\end{eqnarray}
\begin{eqnarray}
F_6 &=& {\textstyle \frac16} C \lbrace S_T^6
+ 6 S_T^2 \mid S^2 \mid ^2 + 24 S_T^2 \mid m S \mid ^2
+ 12[ m^2 S^2 (S^\ast)^2 + c.c.]
\nonumber \\
&+& 3[ 4 S_T^2 (mS)^2 + S_T^4 S^2 + 8 \mid mS \mid ^2 S^2
+ \mid S^2 \mid ^2 S^2 + c.c.] \Delta_{2Q,G}
\nonumber \\
&+& 2[ 6 S_T^2 mS^3 + 4 (mS)^3
+ 3 (mS^\ast)S^4 + c.c.] \Delta_{3Q,G}
\nonumber \\
&+& 3[  S_T^2 S^4 + 4 m^2 S^4 + c.c.] \Delta_{4Q,G}
+ [ S^6 + c.c.] \Delta_{6Q,G} \rbrace,
\end{eqnarray}
where $S_T^2 = m^2 + 2 \mid S \mid ^2$.  The coefficients in (6) are given by
$A_q = aT-J_q$, where $J_q$ is the Fourier transform of the exchange integral,
\begin{equation}
J_q = 4J_0 + 2[J_1 cosq + J_2 cos(2q) + J_3 cos(3q)],
\end{equation}
with $q=cQ$.  The commensurate states $\langle 1 \rangle$ and
$\langle 12 \rangle$ are represented here by $q=\pi$ and $q=\frac23 \pi$,
respectively.  The $IC$ phase is determined by the value of $q$ which maximizes
$J_q$.  This is usually the first ordered state which occurs as the
temperature is lowered from the paramagnetic ($P$) phase, with the
N\'eel temperature $T_{N1} = J_q/a$.  Due to the Umklapp terms, the free
energies of each of these ordered states must be evaluated separately.
None of the Umklapp terms contribute to $F$ of the $IC$-phase.
All non-Umklapp terms, as well as those with $n=2$, 4, and 6, must be
included in the free energy for the $\langle 1 \rangle$-phase, whereas
only non-Umklapp terms and those with $n=3$ contribute to the
$\langle 12 \rangle$-phase.  These separate free energies must be minimized
as a function of the relevant variables in each case: $F_{IC}(m,\mid S\mid
,q)$,
$F_{\langle 1 \rangle}(m,S)$, and $F_{\langle 12 \rangle}(m,\mid S \mid,\phi)$,
where $S=\mid S \mid e^{i \phi}$.  The stable phase as a function of
temperature, magnetic field and exchange parameters, is then determined by
comparing the numerical results for these free energies. For the
$\langle 12 \rangle$-state, the free energy is minimized by a phase angle
$\phi = (2m+1) \pi /3$, where $m$ is an integer.

It is convenient to consider phases which occur as a function of
$J_2/J_1$, with $J_3/J_1$ set to a small value.\cite{yam,selk,hass}
In the absence of a magnetic field,
a number of sequences of phases results from the present model depending
on $J_2$.  With increasing temperature, the following can occur (see I)
$\langle 1 \rangle \rightarrow P$,
$\langle 1 \rangle \rightarrow IC \rightarrow P$,
$\langle 1 \rangle \rightarrow \langle 12 \rangle \rightarrow IC
  \rightarrow P$,
$\langle 12 \rangle \rightarrow IC \rightarrow P$.
The resulting $J_2/J_1 - T$ phase diagram corresponds qualitatively to the
Monte-Carlo simulation results (see Fig. 1 of Ref.\onlinecite{hass}).
(The sequence
$\langle 12 \rangle \rightarrow \langle 1 \rangle \rightarrow IC
  \rightarrow P$
observed in $UNi_2Si_2$ cannot be explained by mean-field theory.)
All transitions except for $IC-P$ are first order.
It is important to note that the $n=3$ Umklapp terms in (7) and (8)
are {\it linear} in $m$.  As a consequence, the spin density of the
$\langle 12 \rangle$ phase has a non-zero uniform component even
in the absence of an applied field.
It is this coupling between $m$ and $S$ which enhances the stablity
of this state in the presence of a magnetic field.

Magnetic phase diagrams were calculated in the manner described
above.  Exchange parameters were set to $J_0=1$, $J_1=-1$, and
$J_3=0.03$.  With these values, phases $\langle 1 \rangle$
and $\langle 12 \rangle$ are degenerate in energy at $T=0$ with\cite{yam,hass}
$J_2$ \hbox{$<$\kern -0.8em\lower 0.8ex\hbox{$\sim$}} 0.35
and the ordering wave vector
in the $IC$ phase is approximately $\frac34 \pi$, as observed
experimentally in some $AT_2X_2$ compounds (such as $UPd_2Si_2$).
There is no temperature or
field dependence of the $IC$ wave vector within the present model.  This
may be included by adding biquadratic exchange (see I).

Typical results shown in Figs. 1 and 2 demonstrate the increasing stability
of the $\langle 12 \rangle$-phase with increasing field.
(This feature is also observed in $UNi_2Si_2$.\cite{reb}) The phase
diagram of Fig. 1 corresponds qualitatively to experimental results
on $UPd_2Si_2$\cite{collins}.  Only relatively low field strengths were
available in this experiment and the merging of the
$\langle 12 \rangle - IC$ and $IC-P$ boundary lines was not observed,
although there is clear indication of this tendency.
For smaller values of $J_2/J_1$, the
$\langle 12 \rangle$-phase may not occur at available field strengths.
At even smaller values of this parameter, the $IC$-phase also disappears.
For larger $J_2/J_1$, only the $\langle 12 \rangle$-state, and perhaps
a small region of the $IC$-state, will occur in the phase diagram.
At sufficiently large values of this parameter, a period-4 state
is stabilized.\cite{yam,hass}

In conclusion, a simple mean-field model based on a Landau-type
free energy derived from molecular-field theory has been shown to
capture the essential features of the ANNNI model in a magnetic
field.  It describes many of the sequences of magnetically ordered
states which occur in axial $AT_2X_2$ compounds.  In particular,
the magnetic phase diagram of $UPd_2Si_2$ is well described by this
model.  It is hoped that this work will serve to stimulate further
experimental investigation of magnetic field effects in this class
of materials.

\acknowledgements
The author is grateful to M.F. Collins for making available preprints
of Refs. 3 and 4 prior to publication.
This work was supported by NSERC of Canada and FCAR du Qu\'ebec.
%


\begin{figure}
\caption{Magnetic phase diagram with exchange parameters
$J_0=1$, $J_1=-1$, $J_2=-0.30$, and $J_3=0.03$.  Solid and broken curves
represent first-order and continuous transitions, respectively.
}
\label{fig1}
\end{figure}

\begin{figure}
\caption{As in Fig. 1, with $J_2=-0.35$.}
\label{fig2}
\end{figure}

\end{document}